\documentstyle[aps,psfig,twocolumn]{revtex}
\begin{document}
\twocolumn[
\hsize\textwidth\columnwidth\hsize\csname @twocolumnfalse\endcsname

\title{Strain driven charge-ordered state in La$_{0.67}$Ca$_{0.33}$MnO$_3$}
\author{Amlan Biswas, M. Rajeswari, R. C. Srivastava,
T. Venkatesan,
and R. L. Greene}
\address{Center for Superconductivity Research, University of Maryland,
College Park, MD-20742}
\author{Q. Lu and A. L. de Lozanne}
\address{Department of Physics, University of Texas, Austin, TX-78712}
\author{A. J. Millis}
\address{Center for Materials Theory,
Department of Physics and Astronomy,
Rutgers University, Piscataway, NJ-08854}
\date{\today}
\maketitle
\tightenlines
\widetext
\advance\leftskip by 57pt
\advance\rightskip by 57pt

\begin{abstract}
We present evidence for the
coexistence of ferromagnetic metallic and charge ordered
insulating phases in strained thin films of
La$_{0.67}$Ca$_{0.33}$MnO$_3$ at low temperatures. 
Such a phase separated state
is confirmed using low temperature magnetic
force microscopy and magnetotransport measurements. 
This phase separated state is not observed in the bulk
form of this compound and is caused by the 
structural inhomogeneities due to the
non-uniform distribution of strain in the film. The strain
weakens the low temperature ferromagnetic metallic state and a
charge ordered insulator is formed at the high strain regions.
The slow
dynamics of the transport properties of the mixed phase is
illustrated by measurements of the long time scale relaxation
of the electrical resistance.

\end{abstract}
]
\narrowtext
\tightenlines

\section{Introduction}
The hole-doped perovskite manganites are of great current interest
for the range of novel properties they display and for the possibility
they offer of new devices. Thin films are important both for 
fundamental studies and for potential applications, but the relation 
between film properties and those of bulk materials is not well
understood. In this paper we show that strain associated with a lattice
mismatched substrate can cause new electronic behavior, not found in 
bulk materials of the same chemical composition.
In hole-doped manganites with the chemical formula
$RE_{1-x}AE_x$MnO$_3$ ($RE$ is a trivalent rare earth ion and
$AE$ is divalent alkaline earth ion)
a complex interplay of the charge, lattice,
spin and orbital degrees of freedom leads to various novel
properties. Of special interest are 
the metal-insulator (MI) transition at the ferromagnetic Curie
temperature ($T_C$) and the phenomenon of charge ordering. These
transitions occur at a wide range of temperatures depending on the
compound i.e. factors like the hole doping concentration $x$,
the average A-site cation radius
$<r_A>$ (A-site refers to the $A$BO$_3$ perovskite structure, where the
A-site is occupied by the RE or AE ion and the B-site is occupied by
the Mn ions),
and the variance in the
radii of the A-site cations $<\sigma_A>$
all of which effectively perturb one or
more of the degrees of freedom mentioned above ~\cite{attfield,cnr}.
Introducing such perturbations
profoundly affects the stability of the ground state and 
gives more insight about the energy considerations for these
transitions.

Beyond the pure phase behavior,
the observation of electronic phase separation
in these materials has generated considerable interest ~\cite{cheong,aarts}.
It has been shown that for a wide range of temperatures there is a
coexistence of ferromagnetic metallic (FMM) and charge ordered
insulating (COI) phases in some of 
these materials. This phenomenon has been
observed for different values of $x$, $<r_A>$, and $<\sigma_A>$.
The low temperature FMM state for $x \sim 0.3$ and
the phenomenon of 
charge ordering (CO) for $x=0.5$ is observed in a variety of
manganite systems and has been studied extensively, both
experimentally and theoretically.
However, as $<r_A>$ is reduced
below about 1.18 \AA, the FMM state disappears and the paramagnetic
insulating state goes directly to the charge ordered insulating
state at low temperatures as observed in the compound
Pr$_{1-x}$Ca$_x$MnO$_3$. This CO state can be made progressively
unstable by replacing the Pr ion with the larger La ion. 
A focus of particular attention is the compound
(La$_{1-y}$Pr$_y$)$_{5/8}$Ca$_{3/8}$MnO$_3$. For y=1, the COI
phase is observed and for y=0 the FMM phase is observed, at low
temperatures. For intermediate values a y, multiphase behavior
is observed at low temperatures characterized by the history
dependence and slow time relaxation of the resistivity. 

In these
compounds the COI and the FMM states have very similar energies,
resulting in the two-phase coexistence of FMM and COI regions 
~\cite{cheong}.
The similarity of the energies of the FMM and COI phases leads
to a
sensitive balance between the FMM and COI regions, which can be
tilted in favor of the FMM phase by the application of a magnetic
field or external hydrostatic pressure and towards the COI phase
by an increase in the structural distortion away from the ideal
cubic perovskite lattice. The structural distortion can be increased
by applying internal pressure by the substitution of smaller ions at
the A-site or by applying anisotropic stress ~\cite{attfield,li,amlan1}.

These observations suggest that interesting phenomena may be 
observed in highly strained thin films of manganites.
In a previous work we have shown the existence of a strain-induced
insulating state in biaxially strained films of  
La$_{0.67}$Ca$_{0.33}$MnO$_3$ (LCMO)  and we claimed that this
was caused by the non-uniform distribution of the strain which
leads to a possible two-phase behavior ~\cite{amlan1}.
Here, we present, detailed studies of the magnetic structure, using
low temperature magnetic force microscopy (MFM), and the 
magnetotransport properties of such biaxially strained films of
LCMO that lend strong support to our claims.
While in its bulk form, LCMO is a 
ferromagnetic metal at low temperature with
a pseudo-cubic lattice structure,
we show that the biaxial strain results
in a distortion of the lattice structure of LCMO.
This distortion makes the low temperature FMM state unstable and a
CO state develops instead. Furthermore, we show that
the strain is non-uniformly
distributed, which leads to a two-phase coexistence of FMM and COI regions
very similar to that exhibited by 
(La$_{1-y}$Pr$_y$)$_{5/8}$Ca$_{3/8}$MnO$_3$.
This is confirmed using low temperature MFM measurements.
We present the various novel properties of this
strain driven two-phase state and discuss the implications on the
properties of hole-doped manganites.

\section{Experimental Details}
Thin films of La$_{0.67}$Ca$_{0.33}$MnO$_3$ (LCMO), 150 \AA~ in thickness,
were grown on (001) LaAlO$_3$ (LAO) and (110)
NdGaO$_3$ (NGO) substrates
by pulsed laser deposition (PLD).
There is a
compressive lattice mismatch strain of $\sim$2\% for a film of LCMO
on LAO while on NGO this strain
is negligible (less than 0.1 \%). Details of the film growth are given in
~\cite{amlan1}.
The resistivities were measured by the conventional four-probe
method and the DC magnetization was measured using a SQUID magnetometer.
The lattice parameters were measured using a Siemens D5000 diffractometer
equipped with a four circle goniometer.
The nanostructure of the films were measured using a
Nanoscope III AFM operated in the tapping mode.

The low temperature MFM has been described in detail previously
~\cite{lozanne1,lozanne2}.
It is a frequency-modulated MFM that measures the magnetic force gradient
$F'$(x,y) distribution on a sample. Briefly, a positive feedback
loop forces a piezo-resistive cantilever to vibrate at its resonant
frequency, $\omega_{0}$. A magnetic tip with iron coating is located at
the free end of the cantilever. During an imaging process, the non-uniform
magnetic stray field from the sample will cause this $\omega_{0}$ to
change. A second feedback system prevents $\omega_{0}$ from shifting by
adjusting
the tip-sample distance. The output of this feedback system is used to
construct an image of constant force gradient. This 
low temperature MFM has been used to
study a variety of materials
~\cite{lozanne1,lozanne2,lozanne3,lozanne4,lozanne5}.

The cantilever we used is a commercial PiezoleverTM self-sensing
device. It is a 2K$\Omega$ piezo-resistive cantilever which is
305$\mu$m long, 50 $\mu$m wide and 3 $\mu$m thick [Thermomicroscopes
technical
datasheet]. The cantilever's force constant is k = 1 N/m and its resonant
frequency is 33 KHz. The tip extends about 2 $\mu$m from the
cantilever. The tip was coated with a 60-nm-thick iron and magnetized
along the tip axis, z, using a permanent magnet. The force gradient
was around 4 $\times 10^{-3}$ N/m 
with small changes from one image to another.

\section{Results and Discussion}

\subsection{DC transport and magnetization}

Figure 1 shows the resistivity and magnetization of a 150 \AA~ film
of LCMO on LAO and for comparison, the resistivity of a 150 \AA~
film of LCMO on NGO.
This figure summarizes the effect
of lattice mismatch strain on the transport and magnetic properties.
The film of LCMO on LAO
is insulating but a metallic temperature dependence appears
in an applied field of 8.5 T whereas
a film of the same thickness
on the lattice matched substrate NGO
shows a resistivity behavior very close to that of
bulk LCMO ~\cite{jaime}.
The magnetization ($M$) starts rising around 250 K but this rise is much
slower than what is observed in thicker films of LCMO on LAO
~\cite{jaime}. The inset
shows the $M$ vs. $H$ curve for the film on LAO
at 5 K. The saturation value of
$M$ ($M_{sat}$) is $\sim$ 1.8 $\mu_{B}$ which is about 50 \% of the expected
$M_{sat}$ = 3.67 $\mu_{B}$.
\begin{figure}
\centerline{
\psfig{figure=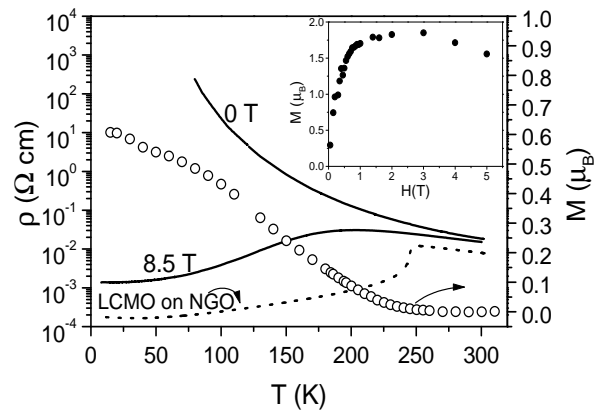,width=8.0cm,height=6.0cm,clip=}
}
\caption{A summary of the transport and magnetic properties of the
150 \AA  film of LCMO on LAO. The resistivity behavior of the film
of LCMO on NGO in zero field is shown for comparison. In zero field
the film on LAO
is insulating at low temperatures. In a field of 8.5 T a metallic state
appears at low temperature. The magnetization
(taken in a field of 1000 G) shows a gradual rise
as the temperature is lowered below $\sim$ 250 K. The inset shows that
the film on LAO has a reduced saturation magnetization of
$\sim 1.8 \mu_B$ at 5 K.}
\end{figure}

From these data we conclude that in zero field most of the volume
of the film on LAO is in an insulating state, and the application
of an 8.5 T field converts a sufficient volume fraction to a 
metallic state to allow metallic conduction. The magnetization
data shows that the $M_{sat}$ is about half of what is expected. This
could be due to: (1) Spin-canting and the out-of-plane
orientation of the spins resulting from the strained growth and (2) phase
\twocolumn[
\hsize\textwidth\columnwidth\hsize\csname @twocolumnfalse\endcsname
\begin{figure}
\centerline{
\psfig{figure=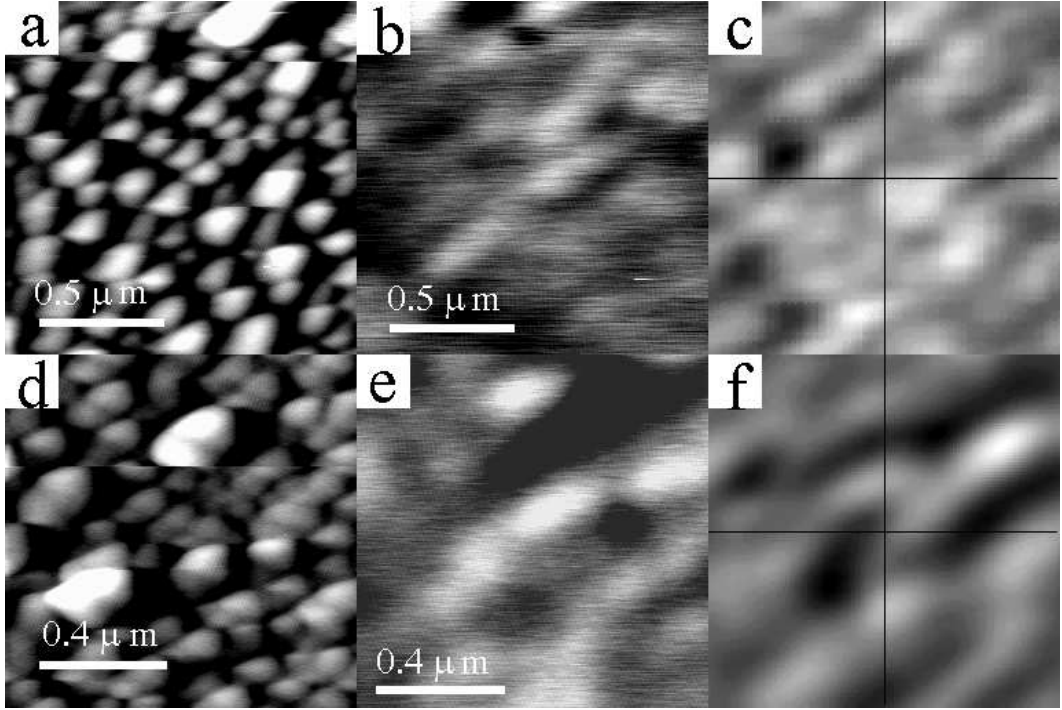,width=14.0cm,height=9.6cm,clip=}
}
\caption{MFM images taken at 80 K of the film of LCMO on LAO. The
images (a) and (d) are topographic images, which are
1.4 $\mu m \times 1.4 \mu$ m and 1.1 $\mu m \times$ 1.1 $\mu m$ 
respectively.
(b) and (e) are the corresponding magnetic images of the same area.
(c) and (f) are the cross correlation functions of the corresponding
topographic and magnetic images. The axes for the displacements
in the x and y directions are marked.
The areas of the cross correlation images are the same as their
respective scan areas. Note the shifts of the regions
of maximum correlation (brighter regions) from the origin in (c) and
(f) (the center of the image is the origin).}
\end{figure}]
separation into FMM and COI phases. Since, from the transport data it is clear
that in zero field a large volume fraction of the film is in the
insulating state [we will argue in following sections that it is a 
charge-ordered insulating state] 
which in manganites is a non-ferromagnetic state, we
claim that the reduction in the $M_{sat}$ is mainly due to the phase
separation into ferromagnetic and non-ferromagnetic phases. In the
following sections magnetotransport and MFM measurements will strengthen
this claim and we will discuss the nature of these two phases.

In an earlier paper we showed using atomic force
microscopy and cross sectional
high resolution transmission electron
microscopy
results that under compressive strain the film grows in the form of
islands and the strain was distributed non-uniformly through the sample
~\cite{amlan1}.
From these observations we propose the following hypothesis.
The origin of the two-phase behavior is the non-uniform strain
produced in the compressively strained film due 
to the island growth mode ~\cite{amlan1,washburn}.
The initial layers of the film grow coherently with the substrate
and are therefore under uniform compressive stress. After a certain
thickness the 3D-islands are nucleated. The edges of these islands are
regions of high strain while the top of the islands are regions of low
strain.
Since, from the magnetization data, it is clear that the low temperature
phase is ferromagnetic albeit with a reduced magnetic moment, it
was suggested
in ref. ~\cite{amlan1} that the top of the 3D islands are regions of
low strain and are FMM at low temperatures. These FMM regions are separated
by the high strain regions at the edges of the islands which are in the
COI phase due to the strain induced lattice distortion. In contrast,
the lattice matched film on NGO grows in the step flow mode and even
150 \AA~ films of LCMO display the properties shown by bulk LCMO and
there is no suggestion of a two-phase behavior ~\cite{amlan1}.
In the remainder of this paper we will present strong evidence 
in support of this hypothesis 
and discuss in detail the reasons and implications. 

\subsection{Low temperature MFM measurements}

To confirm the hypothesis mentioned above, we first obtained images of the
magnetic domain structure in the films of LCMO on LAO
(which shows the island growth mode) and LCMO on NGO (which shows
the step flow growth mode)
using a low
temperature MFM.
Although a calibration of the actual local field 
and its gradient
is difficult from an MFM measurement, we can get important information
about the magnetic structure of the films at a scale
of $\sim$ 500 \AA.
Figure 2 shows the MFM images of the films
of LCMO on LAO. Figure 2a and 2b are 
\twocolumn[
\hsize\textwidth\columnwidth\hsize\csname @twocolumnfalse\endcsname
\begin{figure}
\centerline{
\psfig{figure=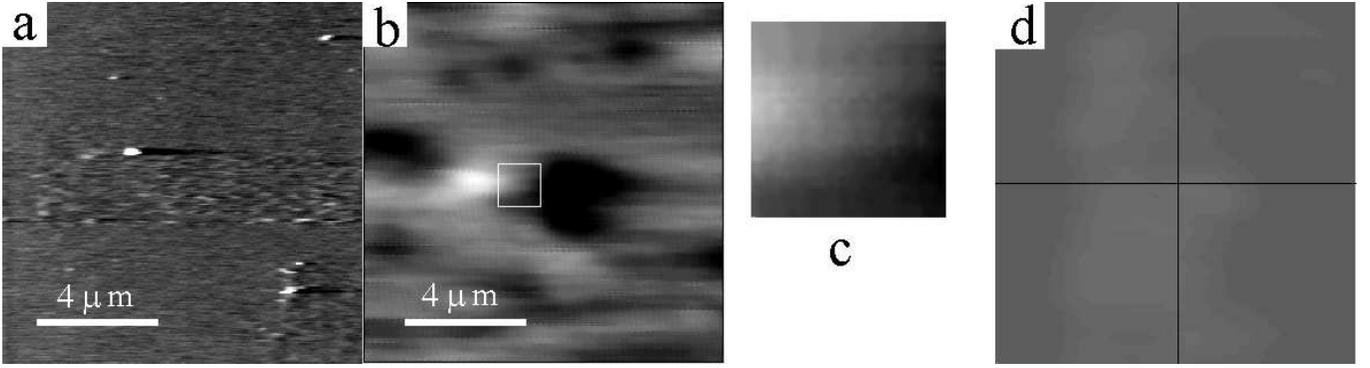,width=18.0cm,height=5.2cm,clip=}
}
\caption{(a) Topographic and (b) magnetic
images of an LCMO film on NGO taken at
80 K. The scan area is 12 $\mu m
\times 12 \mu m$. The square marks an area of 1.4 $\mu m \times 
1.4 \mu m$
which is shown magnified in (c). The domain sizes are much larger
than for the film on LAO. (d) shows the cross-correlation of the
topographic and magnetic images and has the same
area as the scan size. The
featurelessness of the cross correlation
shows that the magnetic structure
for this film grown under negligible strain is uncorrelated
to the topography.}
\end{figure}]
topography and magnetic images
respectively, taken over the same area and figure 2d and 2e are
topography and magnetic images taken over another area.
First we make some qualitative comments about
the images.
The magnetic images clearly 
shows the presence of ferromagnetic (FM) regions in the film on LAO
which shows that at low temperatures a significant part of the
film is in the FMM state although the film is insulating down to the
lowest temperatures. Also, the domain sizes are of the same order
as the island sizes. 
To analyze the images quantitatively, in
figure 2c and 2f we show the 2D cross 
correlation images of the topographic
and magnetic images. 2D cross correlation is the usual method
for 2D pattern recognition. The cross correlation
images have their origin (zero displacement) at the
center of the image and their areas are the same as the corresponding
topographic and magnetic images. The positions of high correlation
in the images (brighter regions) mark the relative displacements at
which the topography and the magnetic images are highly
correlated. From figure 2c and 2f we can see that
there is a high correlation 
between the topography and magnetic images which, for
one, shows that the magnetism of the film is 
linked to the topography. It also shows that
the maximum correlation occurs at points slightly 
offset from the origin. Which means that the highest correlation
between figures 2a and 2b and figures 2d and 2e happens when
the topography and magnetic images are displaced relative to
each other.
This is understandable since the MFM image picks up 
the magnetic force gradient which is 
maximum at the edges of the islands [the change in magnetization is
expected to be maximum at the edges of the islands since the strain changes
rapidly at these locations]. 
The maxima in the cross correlation are hence
shifted from the origin by a distance of the order of the 
average island size. 
The images of the film on NGO are shown for
comparison in figure 3.
The domain sizes are much larger ($\sim 5-10 \mu$m) than the film
on LAO. This is the typical domain size observed in the FMM state of
manganites ~\cite{lozanne1}. There is no correlation between the
topography and the magnetism of the film as seen in figure 3d.

Although these results are limited by the resolution of the MFM
and the fact that we cannot get a direct measure of the actual
local field, 
the results shown above lend strong support to our
claim that the magnetic and topographic structure of the film on 
LAO are linked to each other. This is due to the variation 
of the local
strain in the film growing under compressive strain
in the form of islands. The non-uniform strain results in the 
coexistence of FM regions (in the areas of low strain) and non-FM regions
(in the areas of high strain). In the next subsection we discuss the 
properties of the non-FM regions in detail.
 
\begin{figure}
\centerline{
\psfig{figure=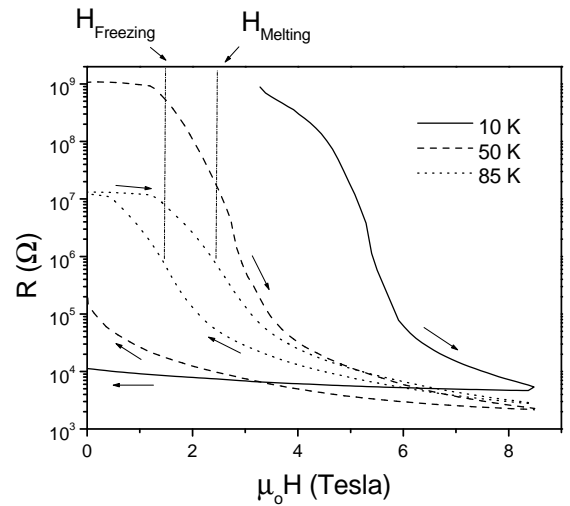,width=8.0cm,height=7.0cm,clip=}
}
\caption{$R$ vs. $H$ curves for the film of LCMO on LAO for three
temperatures. The
melting and freezing fields are marked.}
\end{figure}

\subsection{The $T-H$ phase diagram}

Two points are confirmed by the discussion above, 1) the local
magnetism of the non-uniformly strained
LCMO film on LAO is linked to the microstructure and 
hence to the
distribution of strain and 2) the film has
a significant
portion of FM regions at low temperature,
which
accounts for the magnetic moment of $\sim 1.8\mu_B$ observed at low
temperatures. Yet the film is insulating and has a large MR at low
temperatures which is due to the non-FM
regions mentioned earlier. 
We now investigate this insulating state further.
\begin{figure}
\centerline{
\psfig{figure=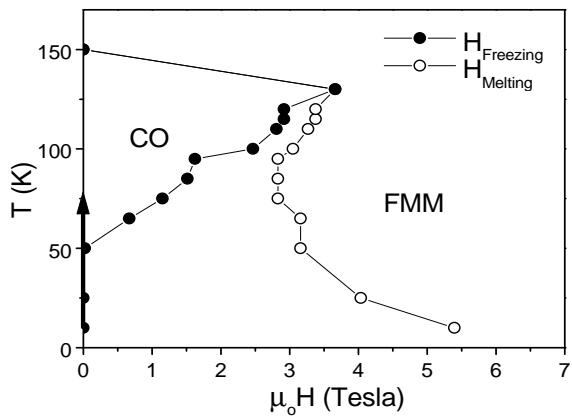,width=8.0cm,height=6.0cm,clip=}
}
\caption{The $T-H$ phase diagram of the film of LCMO on LAO. The
phase diagram closely resembles that for
Pr$_{1-x}$Ca$_x$MnO$_3$ ($x \sim 0.33$). The arrow on the y-axis
shows the path in the phase diagram which was followed to get the
$R$ vs. $T$ curve shown in figure 6.}
\end{figure}
Figure 4 shows the $\rho$ vs. $H$ behavior of the film at low
temperatures. In each case the sample was zero field cooled to that
particular temperature. The magnetic field was then increased and then
reduced to zero. It shows that there is a sharp transition to a low
resistance state above a certain magnetic field and this transition
is associated with a significant hysteresis. This behavior is similar
to the phenomenon of melting of a charge ordered state by a magnetic
field which is a first order transition and is associated with a large
temperature dependent hysteresis ~\cite{tokura4}.
A set of $R$ vs. $H$
curves taken at different temperatures has been
used to construct a phase
diagram in the $T-H$ plane. Such a phase diagram for the 150 \AA~
film 
on LAO
is shown in figure 5. $H_{Freezing}$ and $H_{Melting}$ were
determined from the $R ~vs. H$ curves as the field values at which 
$dR/dH$ was maximum (Fig. 4).
These data show that the insulating state
at low temperatures displays a behavior just like
bulk compounds in the charge ordered state ~\cite{tokura4,tokura3}.
Note that
in bulk La$_{0.67}$Ca$_{0.33}$MnO$_3$ there is no charge ordered
state at low temperature but a low
temperature FMM state. Therefore we conclude
that the regions of high strain (near the edges of the islands)
have a distortion built into the lattice structure due to the substrate
induced strain which leads to the destabilization of the FMM state
at low temperatures and the charge ordered state takes over.
Furthermore, the phase
diagram shows that for $T < 50$K the insulating state does not
recover even when the field is reduced to zero. Such behavior is
similar to that observed in compounds with $<r_A> \le 1.18$ \AA~
such as Pr$_{1-x}$Ca$_x$MnO$_3$.
It has been shown for uniformly strained films that by tuning the
$c/a$ ratio, the magnetic and electronic phases can be controlled
~\cite{tokura2}.
The data for thin films of La$_{1-x}$Sr$_x$MnO$_3$ grown on different
substrates with different amounts of strain (and hence different $c/a$
ratios) were compared to the data for Nd$_{1-x}$Sr$_x$MnO$_3$ for
different $x$ (which varies the $c/a$ ratio in this case). In our films
we have shown that the strain is non-uniform due to the island like
growth mode which suggests that the
$c/a$ ratio varies over the film. The average $c/a$ ratio is $\sim$ 1.04 
(as determined from the lattice constant measurements) but from the MFM
and magnetization results it is clear that some parts of the film
(the top of the islands) are relatively strain free i.e. $c/a \approx
1$. The LCMO film on LAO has therefore been distorted locally resulting
in a variation of the $c/a$ ratio over the film and hence a variation
of the magnetic and electronic properties.

It has been shown by D\"{o}rr {\em et al.} that in 
oxygen deficient films of LCMO a phase separation between FMM
and an antiferromagnetic insulating phase is observed ~\cite{dorr}. 
This is
due to reduced Mn$^{4+}$ concentration in the sample. Even in 
our samples
oxygen annealing of the film on LAO results in the film becoming
FMM at low temperatures ~\cite{amlan1}. However, we argue in the following
that the two-phase
behavior we observe is strain driven and not due to variations
in stoichiometry and the oxygen annealing results mainly 
in a reduction
of this strain. The basis for our argument is the $T-H$ phase 
diagram in
fig. 5. In the bulk La$_{1-x}$Ca$_x$MnO$_3$, there is no value of
$x$ for which there is such a $T-H$ phase diagram. The most striking
evidence is the fact that at 10 K there is a drop of about 5 orders
of magnitude in the resistance at a field of about 6 T and the
insulating state is not recovered when the field is removed (fig. 4). 
Also for the CO state seen for $x \ge$ 0.5 in bulk 
La$_{1-x}$Ca$_x$MnO$_3$,
the melting fields are about 14 T and higher
~\cite{xiao,book}. Therefore, this strain driven phase of LCMO
cannot be explained as being due to a variation in stoichiometry.
In fact this state is similar to the charge ordered state in the
Pr$_{1-x}$Ca$_x$MnO$_3$ system for $x \sim$ 0.3, as can be seen from the
phase diagram in figure 5 ~\cite{tokura3}.
Therefore this two-phase state
comprising FMM and COI regions
obtained by applying biaxial strain is a new, strain driven, state
of LCMO. We are further investigating the additional role of
oxygen deficiency on such a strain driven COI state by systematically
reducing the oxygen content of our films on LAO.

\subsection{Metastable states and long time scale dynamics}

From figure 5 we observe that the width of the hysteresis in the
$R$ vs. $H$ curves increases as the temperature is lowered. This
is due to the decrease of the effect of thermal fluctuations on the
first order transition ~\cite{tokura4}.
As mentioned earlier, for $T < 50$K the metallic state obtained
on melting the CO state by a magnetic field is retained even after the
magnetic field is removed. This is a metastable state and the high
resistance state is recovered when the temperature is raised above 50 K
\begin{figure}
\centerline{
\psfig{figure=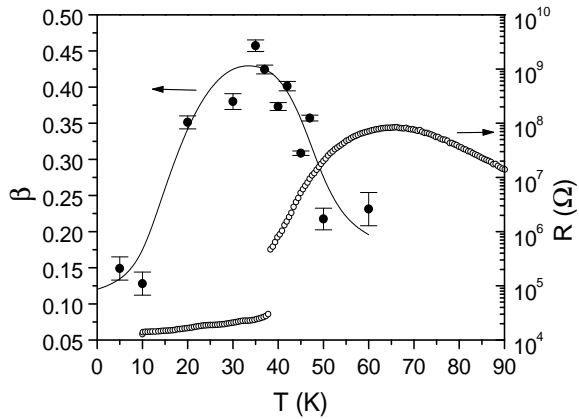,width=8.0cm,height=6.0cm,clip=}
}
\caption{The values of $\beta$ as a function of temperature. The
solid line is a guide to the eye. The
behavior of the resistance along the path marked in figure 5 is
also shown. There is large jump in the resistance around 40 K and the
insulating state is recovered near 65 K.}
\end{figure}
as shown in figure 6.
This association
of the two-phase coexistence with the local
structural distortions results in slow
dynamics which we have measured as the slow relaxation of the resistance of
a metastable state. We obtained the metastable state as follows. We first
cooled down the sample in zero field and applied a magnetic field of 8.5 T
at 5 K. A metallic state was thus obtained. The field was then removed but
the film was still trapped in the metastable metallic state.
We then measured the resistance as function
of time of this metastable state. The resistance increases as a function
of time. The relaxation of the resistance can be fitted with a stretched
exponential form $R(t) = R_0 - R_1 exp[-(t/\tau)^\beta]$.
The stretched exponential form of the relaxation has been observed in
numerous systems but the exact physical significance of the fit
parameters $\beta$ and $\tau$ is not fully understood. 
In particular, such long time scale relaxation of the resistance
has been observed in bulk charge-ordered systems like
La$_{0.5}$Ca$_{0.5}$MnO$_3$ and La$_{5/8-y}$Pr$_y$Ca$_{3/8}$MnO$_3$ 
~\cite{vera,cheong2}. One
conclusion which can be drawn from these observations is that there
are metastable states which are separated by energy
barriers with a wide
distribution of energies which leads to a distribution of relaxation times
and the stretched exponential form of relaxation. The time constant
$\tau$ has been observed to be thermally activated i.e. of the form
$\tau = \tau_0 exp(E_{\tau}/k_BT)$.
$E_{\tau}$ is related to the distribution of physical activation
energies ~\cite{benatar}. A variation of $E_{\tau}$ suggests that
the activation energies in the system have been changed.
For the film of LCMO on the LAO, $E_{\tau} \sim$ 96$\pm$5 K
(figure 7). It is
interesting to compare the value of $E_{\tau}$ with the field required
to melt the charge ordered state at low temperatures ($H_{M0}$).
Ref. ~\cite{cheong2} gives the values of $E_{\tau}$
for the two compounds La$_{0.5}$Ca$_{0.5}$MnO$_3$ and
La$_{5/8-y}$Pr$_y$Ca$_{3/8}$MnO$_3$ ($y = 0.375$). The inset of
figure 7 shows a plot of $E_{\tau}$ vs. $H_{M0}$ which shows that
the two are linearly related. This analysis of the slow relaxation
of the resistance shows that the strength of the charge ordered state
depends on the distribution of activation energies between the metastable
\begin{figure}
\centerline{
\psfig{figure=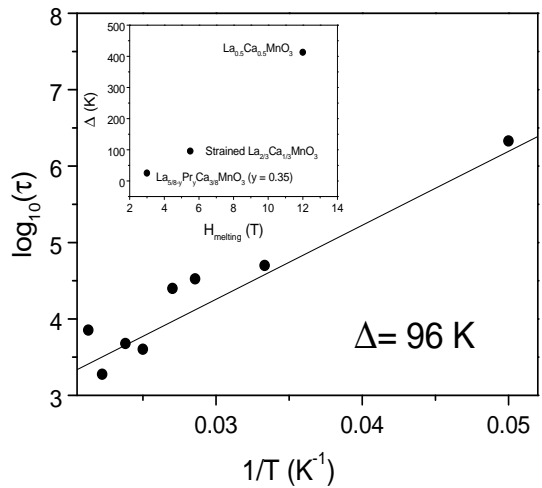,width=8.0cm,height=7.0cm,clip=}
}
\caption{The $1/T$ vs. log $\tau$ plot. The value of $E_{\tau}$ is
96 $\pm$ 5 K. The inset shows a plot of $E_{\tau}$ vs. $H_{melting}$
for three materials. The first and third points are taken from
ref. 21.}
\end{figure}
states. This could presumably be due to the activation energy barriers
being associated with motion of domain walls in the phase separated
regime, which suggests that the field driven transition is perhaps 
not a 
thermodynamic transition.
The physical significance of the value of $\beta$ is less clear. The
only comment which can be made is that
near the temperature where the metastable, low resistance
state makes the transition to the higher resistance state, the $\beta$
value reaches up to $\sim$ 0.5. This indicates that the distribution
of the energies of the barriers is the broadest near the transition
~\cite{benatar}. Similar values of $\beta$ ($0.34 < \beta < 0.46$)
were also obtained from
relaxation measurements in La$_{5/8-y}$Pr$_y$Ca$_{3/8}$MnO$_3$ 
($y = 0.375$) ~\cite{cheong2}.

\section{Summary} 

In this paper we have shown that: (1) Film growth under biaxial
compressive strain readily leads to a 3D island growth mode which leads
to a non-uniform distribution of the strain. This leads to pockets of
high distortion (near the edges of the islands) and regions where the
strain is very low (top of the islands). (2) For a
system like La$_{0.67}$Ca$_{0.33}$MnO$_3$ which is cubic at low
temperatures, the non-uniform strain leads to the ratio $c/a \ne 1$
in the high strain regions and $c/a \sim 1$ in the low strain regions.
(3) The low strain regions are FM at low temperatures and contribute
to the saturation magnetization of 1.8 $\mu_B$. (4) The high strain
regions are in the CO state 
at low temperatures and result in the insulating
behavior of the film in zero field. The magnetoresistance behavior
of the film shows that the strain drives the LCMO film to a state
similar to the low temperature state observed in
Pr$_{1-x}$Ca$_x$MnO$_3$ ($x \sim 0.33$). (5) The two-phase behavior
of the film is coupled to the localized structural distortions and
this leads to slow dynamics in the transport properties of the film.
There is a distribution of the activation energies separating the
metastable states and this distribution of energies determines the
strength of the COI state.
It has been shown in ~\cite{ahn} that there is a strong coupling between
the lattice , Mn $e_g$ orbital state, and the exchange interaction.
When we perturb the lattice by the non-uniform strain it results in
inhomogeneities in the exchange and hopping interactions. Near first
order transitions such disorder leads to large scale two-phase
coexistence ~\cite{dagotto}. This disorder can also be introduced by
chemical substitution ~\cite{cheong2}. Disorder is therefore an 
essential ingredient for the occurence of large scale phase separation.

\section{Conclusions}

The most important claim that we make is that there is a strain
driven COI state in La$_{0.67}$Ca$_{0.33}$MnO$_3$ which is similar
to the small $<r_A>$ compound Pr$_{0.67}$Ca$_{0.33}$MnO$_3$. We 
support this claim with MFM and magnetotransport data. These are
not ``confirmatory tests" for detection of a charge ordered state like
electron diffraction ~\cite{cheong}. We are still trying to devise a 
method to do perform such a confirmatory test on these samples.
Nevertheless, in this paper we have presented compelling evidence
to support our claims. 

The results shown above elucidate the instability of the FMM(COI) phases
towards the COI(FMM) phases on the application of perturbation like
a magnetic field, external or internal stress etc. At certain values of
these external parameters there are first order transitions between these
two phases. Near these first order transitions, the defects in
the sample stabilize the observed large scale
phase separation in hole-doped manganites. Such effects are therefore
intrinsic to the behavior of manganites and have to accounted for when
explanations of the properties of manganites are attempted.

\acknowledgements

This work was partially supported by the MRSEC program
of the NSF at the University of Maryland, College Park (Grant No. 
DMR-0080008).

\end{document}